\documentclass[prl,twocolumn,superscriptaddress]{revtex4}

\usepackage{amsmath}

\usepackage{graphicx}
\usepackage{times}
\usepackage{epsfig}

\begin{document}

 \title{Non-equilibrium transport through a point contact in the
$\nu=5/2$ non-Abelian quantum Hall state}

\author{Adrian Feiguin}
\affiliation{Condensed Matter Theory Center, Department of Physics, University of
Maryland, College Park, MD 20742, USA}
\affiliation{Microsoft Research, Station Q, CNSI Building,
University of California, Santa Barbara, California 93106, USA}
\author{Paul Fendley}
\affiliation{Department of Physics, University of Virginia,
Charlottesville, VA 22904, USA}
\author{Matthew P.A. Fisher}
\affiliation{Microsoft Research, Station Q, CNSI Building,
University of California, Santa Barbara, California 93106, USA}
\affiliation{Department of Physics, University of California,
Santa Barbara, California, 93106, USA}
\author{Chetan Nayak}
\affiliation{Microsoft Research, Station Q, CNSI Building,
University of California, Santa Barbara, California 93106, USA}
\affiliation{Department of Physics, University of California,
Santa Barbara, California, 93106, USA}

\begin{abstract}
We analyze charge-$e/4$ quasiparticle tunneling between
the edges of a point contact in a non-Abelian model of the
$\nu=5/2$ quantum Hall state in the presence of a {\it finite
voltage difference} using the time-dependent density-matrix
renormalization group (DMRG) method. We confirm that, as
the voltage decreases, the system is broken into two pieces.
In the limits of small and large voltage, we recover the results
expected from perturbation theory about the infrared and
ultraviolet fixed points. We test our methods by finding the
analogous non-equilibrium current through a point contact
at $\nu=1/3$.
\end{abstract}

\maketitle


\paragraph{Introduction--}
The $\nu=5/2$ fractional quantum Hall state
\cite{Willet87,Pan99,Eisenstein02} is suspected to be a non-Abelian
topological state. While theoretical evidence has been steadily
accumulating over the years, there has been little experimental
evidence -- until now. Several transport measurements utilizing a
point contact, one a low-frequency noise (`shot noise') measurement
\cite{Dolev08}, the other the tunneling current \cite{Radu08},
indicate that the smallest quasiparticle charge at this plateau is
$e/4$. This is consistent with two non-Abelian models of the
$\nu=5/2$ fractional quantum Hall state
\cite{Moore91,Nayak96c,Levin07,Lee07}, but it also consistent with an
Abelian model \cite{Halperin83}.

One limit to the success of these 
measurements is that the data is compared only to
lowest-order perturbative calculations valid for small
inter-edge tunneling at the point contact.  However, inter-edge
tunneling in these experiments is not so small, and may have effects
not described with low-order perturbation theory. Therefore, it is
important to compute the expected $I\!-\!V$ curve (and the
low-frequency noise) beyond the perturbative regime.  In so doing, we
follow the crossover from the limit of weak quasiparticle tunneling
across the point contact to the low-temperature, low-voltage regime.

In this paper, we numerically compute the zero-temperature current
through a point contact in the presence of a finite voltage bias in
two fractional quantum Hall states. The first is the Abelian Laughlin
state at $\nu=1/3$, which allows us to check our numerical procedure,
and to confirm that indeed the Bethe ansatz computations
\cite{Fendley95} are applicable out of equilibrium. The second is the
non-Abelian Moore-Read Pfaffian state with $\nu=5/2$ \cite{Moore91}.
We show that in both cases the droplet eventually breaks in two at low
voltages.  In the weak quasiparticle tunneling limit, the current
follows the predicted power law. In the low-voltage limit, the
conductance is obtained and is approached in a manner
consistent with predictions \cite{FFN}.


We use the time-dependent Density Matrix Renormalization Group (DMRG)
method \cite{tdDMRG1,tdDMRG2}, because other approaches have
difficulties. Namely, at finite bias, this non-equilibrium
calculation is not amenable to a Monte Carlo simulation.  (The Monte
Carlo computation in the $\nu=1/3$ case is applicable only to the
linear-response regime \cite{Moon93}.)  As we discuss below, the
problem can be mapped onto resonant tunneling between attractive
Luttinger liquids. Since the `leads' are interacting, a Wilsonian
numerical renormalization group cannot be used, unlike in the Kondo
problem. Finally, when $\nu=5/2$, the model is not integrable, and
thus the Bethe ansatz is not applicable.

\paragraph{Models--}
The edge excitations of the Laughlin state at $\nu=1/3$ are described
by a chiral Luttinger liquid. At a point contact, two edges come into
close proximity so that a charge $e/3$ quasiparticle can be
backscattered from one to the other. 
We describe each edge by a chiral boson $\phi_i$, so
that a charge $e/3$ Laughlin quasiparticle is created by the operator
$e^{i\phi_i/\sqrt{3}}$. The effective Hamiltonian is
\begin{equation}
{\cal H}^{1/3} = 
\sum_{i=1,2}\frac{v_c}{4\pi}\,\int dx ({\partial_x}\phi_{i})^2
  \ + \ t\, e^{i(\phi_1(0)-\phi_2(0))/\sqrt{3}} + \text{h.c} 
\label{eqn:Laughlin-pt-contact}
\end{equation}
The tunneling amplitude has scaling dimension $[t]=2/3$.  Hence, at
zero temperature the backscattere current
${I_B}\sim {t^2} V^{-1/3}$ in the limit of small
tunneling current and large voltages.
In the low voltage limit, perfect backscattering occurs and the Hall
bar effectively breaks in two \cite{Kane92}. Charge transport between
the two halves is due to electron tunneling, so for small $V$,
${I_B}-\frac{1}{3}\,\frac{e^2}{h} V \sim V^5$.  The picture was confirmed
by finding the full crossover from weak to strong backscattering via a
Monte Carlo calculation of the linear-response current at non-zero
temperature \cite{Moon93} and the Bethe ansatz solution for the full
${I_B}(V)$ curve \cite{Fendley95}. 

To find the zero-temperature ${I_B}(V)$ curve numerically, we rewrite this
problem as non-resonant tunneling between two semi-infinite {\it
non-chiral} spinless Luttinger liquids. We define ${\phi_a}(x)$ on the
half-line $x<0$ and ${\phi_b}(x)$ on the half-line $x>0$ as follows:
${\phi_a}(x<0)={\phi_1}(x)+{\phi_1}(-x)$ and
${\phi_b}(x>0)={\phi_2}(x)+{\phi_2}(-x)$. The tunneling term in
(\ref{eqn:Laughlin-pt-contact}) then becomes
\begin{equation}
{\cal H}_{\rm non-res} = t\left({\psi_a^\dagger}(0) {\psi_b}(0) +
\text{h.c.}\right)
\label{massterm}
\end{equation}
where ${\psi_{a,b}}(x)=e^{i{\phi_{a,b}}(x)/2\sqrt{3}}$
are the two Luttinger quasiparticle creation operators (which are
non-local combinations of the Laughlin quasiparticles).
Since these operators have scaling dimension $1/6$, the
Luttinger liquids have $g=3$ in the conventions of Ref.\
\onlinecite{Kane92}; a duality transformation maps this to a $g=1/3$
Luttinger liquid perturbed by a $\delta$-function impurity.  


In our DMRG computations, we use a tight-binding model of spinless
fermions with nearest-neighbor attractive interactions to describe
each Luttinger liquid.  This model is equivalent, under a
Jordan-Wigner transformation, to an XXZ spin chain.  Coupling the two
liquids corresponds to including a link between the sites at the two
ends, as illustrated in figure 1a. Just as charge tunneling violates
charge conservation of the individual edges, coupling the two chains
violates the conservation of the individual magnetizations. We match
parameters by noting that the scaling dimension of the staggered
spin-raising operator in the spin chain, $S^\dagger_x (-1)^x$, is
equal to the Luttinger parameter $g$.  The ferromagnetic XXZ spin chain
anisotropy is then related to $g$ by $J_z/J_\perp = -\cos(\pi/2g)$
\cite{Baxter}. Thus $H_{\rm non-res}$ can be
realized with ${J_z}/{J_\perp}=-\sqrt{3}/2$.

Tunneling through a point contact for $\nu=5/2$ also can be realized
via coupled XXZ chains, by following the bosonization procedure of
\cite{FFN}. We assume that the $N=0$ Landau level (of both spins) is
filled and the $N=1$ Landau level is in the half-filled Moore-Read
Pfaffian state. The former are integer quantum Hall edge modes, and
are the outermost excitations of the system; we ignore them because we
focus on tunneling across the interior of a Hall droplet.  The gapless
chiral theory describing the edge excitations of the Moore-Read state
consists of a free boson (the charge sector) and a free Majorana
fermion (the neutral sector). We study the inter-edge backscattering
of the basic charge $e/4$ quasiparticle at a point contact at $x=0$.
Charge-$e/2$ quasiparticles can also tunnel, as can neutral
quasiparticles, but the latter does not affect the electrical
conductivity and the former is expected to be smaller.
The effects of the latter, as well as the extension to the
anti-Pfaffian and $(3,3,1)$ states, will be
discussed elsewhere \cite{Feiguin08}.

To bosonize this model requires a fairly elaborate computation,
because the charge-$e/4$ quasiparticle has non-abelian
statistics. When the dust settles, the tunneling Hamiltonian can be
written in terms of two bosons $\phi_\rho$ and $\phi_\sigma$, and a
Kondo spin $\vec{S}$. The resulting Hamiltonian is \cite{FFN}
\begin{multline}
\label{eqn:edge-Hamiltonian}
{\cal H}_{5/2} = \int_0^\infty dx \left(\frac{v_c}{2\pi}
\left({\partial_x}{\phi_\rho}\right)^2
+\frac{v_n}{2\pi}\left({\partial_x}{\phi_\sigma}\right)^2\right)\\
+  t \left({S^+}e^{-i{\phi_\sigma}{\hskip -0.03cm}(0)/2}+
{S^-}e^{i{\phi_\sigma}{\hskip -0.03cm}(0)/2}\right) 
\cos\!\left({\phi_\rho}(0)/2\right)\ .
\end{multline}
The tunneling amplitude has scaling dimension $[t]=3/4$.
Therefore, in the limit of small current, ${I_B}\sim {t^2} V^{-1/2}$
at $T=0$; this limit occurs for large voltage.
In the opposite limit, $V\rightarrow 0$, inter-edge tunneling becomes
strong, and ${I_B} \rightarrow \frac{1}{2}\,\frac{e^2}{h}\,V$.
Deviations from this total backscattering vary generically as: ${I_B} -
\frac{1}{2}\,\frac{e^2}{h}\,V \sim V^5$.  However, when the tunneling
of neutral Majorana fermions is neglected, the system flows to the
infrared fixed point along a special direction, so that ${I_B} -
\frac{1}{2}\,\frac{e^2}{h}\,V \sim V^{15}$ \cite{FFN}. (Even though Majorana
fermion tunneling does not directly contribute to charge transport, it
affects the flow into the infrared fixed point.)  In this paper, we
will compute ${I_B}(V)$ for arbitrary $V$.

The Hamiltonian (\ref{eqn:edge-Hamiltonian}) has the form of resonant tunneling
between attractive Luttinger liquids; the reason that the Luttinger liquids are
{\it attractive} ($g>1$) is that the tunneling operator has scaling dimension $1/4$,
which is highly relevant.
We can make the relation to resonant tunneling between Luttinger
liquids more apparent by rewriting (\ref{eqn:edge-Hamiltonian}) in the form
\begin{multline}
\label{eqn:res-level}
{\cal H}_{\rm res} = \int_0^\infty dx\,\frac{v}{2\pi}
\left(\left({\partial_x}{\phi_a}\right)^2
+\left({\partial_x}{\phi_b}\right)^2\right)\\
+ t\, d^\dagger e^{i \phi_a(0)/\sqrt{g}} +
t\, d^\dagger e^{i \phi_b(0)/\sqrt{g}}+ \text{h.c.} 
\end{multline}
where ${S^-}=d$, ${S^+}=d^\dagger$ annihilate/create
a particle on the resonant level. The Luttinger coupling is $g=2$ with
$\phi_{a/b} = \frac{1}{\sqrt{2}} ( \phi_\sigma \pm \phi_\rho)$.

As with the earlier case, we utilize two semi-infinite ferromagnetic
XXZ chains for the two Luttinger liquids. Here, the two chains couple
through the resonant level, which in the lattice model corresponds to
adding an extra site, as illustrated in figure 1b. Since $g=2$ here,
we have ${J_z}/{J_\perp}=-1/\sqrt{2}$.


We apply a potential difference $V$ between the two leads, or
potential differences $\pm V/2$ between lead $a/b$ and the resonant
level.  This modifies the tunneling term in (\ref{eqn:res-level}) to
$t\, d^\dagger e^{i {\phi_a}/\sqrt{g}} e^{ieVt/2}+ t\, d^\dagger e^{i
{\phi_b}/\sqrt{g}} e^{-ieVt/2} + \text{h.c.}$.  However, in the
original $5/2$ point contact problem, the tunneling term transfers
charge $e/4$ between the two edges, so a potential difference $V$
between the edges modifies the tunneling term in
(\ref{eqn:edge-Hamiltonian}) to $t \left({S^+}e^{-i{\phi_\sigma}/2}+
{S^-}e^{i{\phi_\sigma}/2}\right) \cos\!\left({\phi_\rho}(0)/2 +
eVt/4\right)$.  Because the current is proportional to the charge
squared, the relation between the current in the original
$5/2$ point contact problem and in the Luttinger liquid resonant
tunneling problem is ${I_B}^{\rm MR} = \left({1}/{2}\right)^2 I_{\rm
res}$.  Likewise in the non-resonant case appropriate for $\nu=1/3$,
we have
${I_B}^{1/3}=(1/3)^2 I_{\rm
non-res}$.

\begin{centering}
\begin{figure}
\epsfig {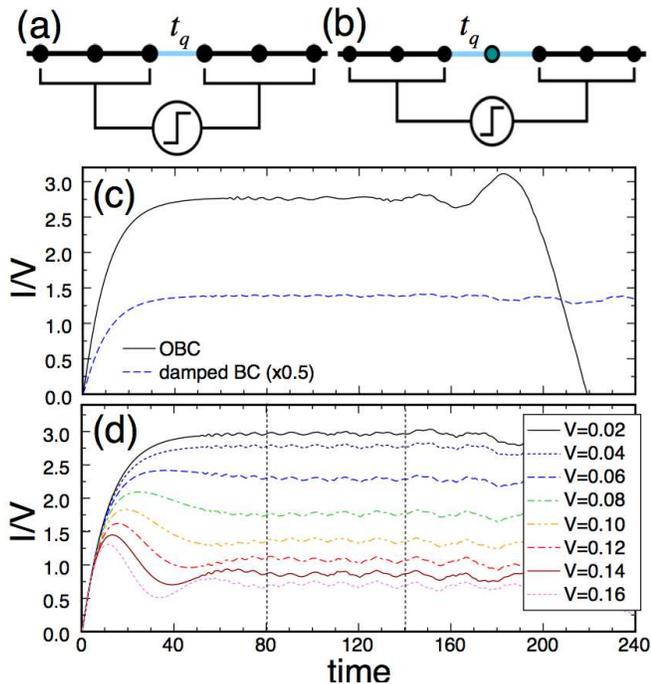}
\caption{ (a) Junction model used to study the current though a
quantum point contact at $\nu=1/3$, consisting of a weak link with
connecting two interacting leads. (b) Quantum dot, or resonant level,
system used for $\nu=5/2$. (c) Current through a
junction after a step bias is applied. We show results for $L=120$,
$t_q=0.1$, $V=0.04$, and different boundary conditions. (d)
Time-dependent current though the same system, with damped boundary
conditions, and different values of the bias $V$. Time is measured in
units of the hopping $t_q$.  } \label{fig1}
\end{figure}
\end{centering}

\paragraph{Time-Dependent DMRG--}


We find the $I$-$V$ curves of the non-resonant and resonant tunneling
problems using the time-dependent DMRG method. Two
fundamental aspects of our calculation make it particularly
unsuited to more conventional techniques such as Wilson's numerical
renormalization group: (i) the conducting leads are interacting
Luttinger liquids and
(ii) we are interested in the non-linear regime, {\it i.e.} large
voltage bias.  In 1D metallic systems, correlations can drastically
affect the density of states and transport properties. In particular,
repulsive interactions suppress charge transport, while attraction
``heals'' the system, enhancing the conductance.  The time-dependent
DMRG is well suited for our systems, because it allows one to
seamlessly incorporate interactions into the leads, and is not
restricted to the linear response regime.

Our technique consists of evaluating the time dependence of the
current though the weak link or quantum dot, after a voltage bias is
applied \cite{Khaled06,LuisDaSilva08}.  In a first step, the ground
state is calculated using the conventional DMRG technique. Then, the
system is quenched: by applying a shift in the chemical potential
$\delta \mu_L = V/2$, $\delta \mu_R = -V/2$ to the left and right
leads, respectively. The resulting non-equilibrium system is evolved
in time by solving the time-dependent Schr\"odinger equation. As a
response to the quench, a current starts flowing through the system.
Typically, the current grows and a transient is observed in the
beginning, followed by oscillations that tends to stabilize at a
constant value, corresponding to the steady state
\cite{Meir93,Jauho94}.  Since the leads used in the calculation are
finite, a reversing of the current is observed after the wave packet
reaches the boundaries and is reflected back. This determines a time
scale in which we expect the current to stabilize at a plateau
value. As we show here, depending on the choice of parameters in the
model, it is sometimes necessary to study large systems in order to
achieve a steady state. In some cases, when the transient region is
large, and the system too small, this is hard to attain.  In order to
improve the behavior of the system, we used long leads and damped
boundary conditions, by exponentially decreasing the coefficients in
the Hamiltonian toward the end of the chains. As a result, removing a
particle from these regions becomes energetically costly, and they
effectively behave as reservoirs. As a consequence, the charge becomes
trapped and accumulates without getting reflected, leading to longer
plateaus\cite{Bohr06,Khaled06}.

\begin{centering}
\begin{figure}
\epsfig {file=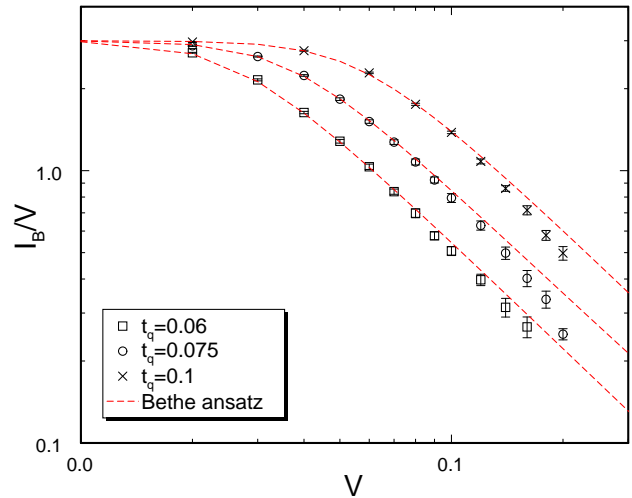,width=85mm,angle=0}
\caption{ ${I_B}$-$V$ characteristics of quasi-particle tunneling
between $\nu=1/3$ quantum Hall edges, as
modelled by a junction system modeling, for
different values of the tunneling amplitude $t_q$, obtained using the
time-dependent DMRG.  Error bars correspond to the errors in averaging
the current over an interval of time of the order of $40$, in unit of
the hopping, and system sizes up to $L=160$.  Lines are fits to the
data using the exact Bethe ansatz solution.  } \label{1-3}
\end{figure}
\end{centering}

\paragraph{Results--}
The tunneling problem between $\nu=1/3$ edges reduces to studying the
current though a weak link, or junction, connecting two interacting
spinless leads,
while the problem at $\nu=5/2$ corresponds to a resonant level, or
quantum dot, as seen in Figs.\ref{fig1}(a) and (b),
respectively. These two models were studied in an early formulation of
the time-dependent DMRG method \cite{Cazalilla03}.  The smaller the
inter-lead hopping amplitude $t_q$, the larger is the initial transient in the
current, making it more difficult to reach a steady state in a finite
system. We found it necessary to use long chains, up to $160$
sites. In Fig.\ref{fig1}(c), we show results for $\nu=1/3$, comparing
the behavior of the current in systems with different sizes and
boundary conditions. The damped boundary conditions, while yielding
the same steady current for given values of $t_q$ and bias $V$,
extend the duration of the plateau, allowing one to reach the steady
state in smaller systems.  In Fig.\ref{fig1}(d) we show results for
$t_q=0.1$, and different values of the bias $V$, for a system with
$120$ sites and damped boundary conditions. Typical simulations extend
to times of the order of $300$ in units of the hopping, using a 3rd
order Suzuki-Trotter decomposition of the evolution operator with a
time step $\tau=0.2$ and keeping the truncation error below $10^{-7}$
\cite{tdDMRG3}. The current is averaged over an interval of time, and
the error calculated following the prescription discussed in Refs.\
\cite{Khaled06,LuisDaSilva08}. The Suzuki-Trotter error associated to
the finite time-step was found to be much smaller than the error in
the average.

In Fig.\ \ref{1-3} we show the ${I_B}$-$V$ characteristic curves for the
$\nu=1/3$ case, for different values of $t_q$. At small biases, the
system exhibits a conductance $G_{\rm non-res}=3{e^2}/h$ which
corresponds to $G_{1/3}=\frac{1}{3}{e^2}/h$, as expected from the
arguments of Ref. \cite{Kane92}.  As the bias grows, the system
departs from the linear response regime and crosses over to the
scaling behavior associated with quasiparticle tunneling at the
ultraviolet fixed point, ${I_B}\propto V^{-1/3}$ \cite{Kane92} (note that
we plot ${I_B}/V$ vs. $V$).  As seen in Fig.\ref{1-3}, not only
the asymptotic power law but the full crossover follows the
behavior predicted by the Bethe ansatz solution \cite{Fendley95}.  The
exact Bethe ansatz expression has a free parameter, corresponding to
the tunneling amplitude or $t_q$, which can be used to fit the
numerical results. The agreement is excellent for small $t_q$, up
until the very large biases at which the lattice model no longer
accurately represents the quantum Hall edge due to curvature of the
dispersion.  At large $t_q$, the system remains near the infrared
fixed point up to large biases, and the scaling behavior associated
with quasiparticle tunneling at the ultraviolet fixed point cannot be
observed.

\begin{centering}
\begin{figure}
\epsfig {file=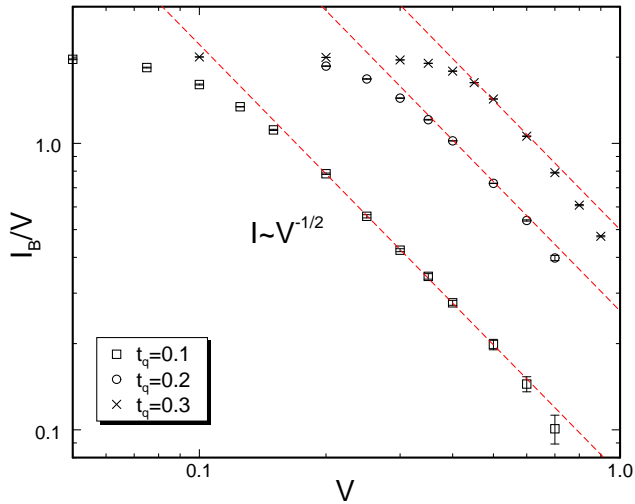,width=85mm,angle=0}
\caption{
${I_B}$-$V$ characteristics of quasi-particle tunneling between $\nu=5/2$ quantum Hall edges, as modelled by a quantum dot system,
for different values of the tunneling amplitude $t_q$,
obtained using the time-dependent DMRG.
Error bars correspond to the errors in averaging the current over an interval of time of the order of $40$, in unit of the hopping, and system sizes up to $L=140$.
Lines are fits to the data using an expression ${I_B}\sim V^{-1/2}$.
} \label{5-2}
\end{figure}
\end{centering}

For the $\nu=5/2$ case, we followed the same 
procedure described above, but using the resonant level system shown in 
Fig.\ref{fig1}(b). The results are depicted in Fig.\ \ref{5-2}. At small
bias, the system exhibits a conductance $G_{\rm res}=2{e^2}/h$
which corresponds to $G_{\rm MR}=\frac{1}{2}{e^2}/h$,
as expected from the arguments of Ref.\ \onlinecite{FFN}.
As the bias is increased for small $t_q$, the system crosses over to the
asymptotic power-law associated with charge-$e/4$ quasiparticle tunneling,
${I_B}\propto V^{-1/2}$. Again, at large bias, the numerical results depart 
from the universal regime, and exhibit the effects of the lattice.
For small $V$, the deviations from $G_{\rm res}=2{e^2}/h$ are too
small to be reliably fit to a power law; this may be an indication
that they are, indeed $\sim V^{15}$. In further work \cite{Feiguin08},
we will investigate whether the addition of an additional marginal operator
will lead to the generic flow into this infrared fixed point with
deviations from the asymptotic value $\sim V^{5}$.

\paragraph{Discussion.}

These results clearly demonstrate that weak inter-edge quasiparticle
tunneling causes a Moore-Read quantum Hall droplet to split into two droplets
which are coupled through weak electron hopping, as
predicted in Ref.\ \onlinecite{FFN}.
They further enable us to access the crossover regime of
intermediate biases where we find quantitative deviations
from power-law behavior. Further work will compute \cite{Feiguin08} the
analogous ${I_B}$-$V$ curves for the anti-Pfaffian \cite{Levin07,Lee07}
and $(3,3,1)$ states \cite{Halperin83} and compare all three to
experimental measurements \cite{Radu08}.
Such a comparison could pave the way to correctly identifying
the $\nu=5/2$ quantum Hall state.

\bigskip 

This work was partially supported by the NSF under
grants DMR/MSPA-0704666 (PF) and
DMR-0529399 (MPAF).
We thank S. Das Sarma and U. Schollw\"ock
for discussions.

\end{document}